\documentclass[twocolumn,floatfix,superscriptaddress,amsmath,aps,prl,longbibliography]{revtex4-2}
\usepackage[utf8]{inputenc}
\usepackage[final]{graphicx}
\usepackage{bm}
\usepackage{amsfonts}
\usepackage{amsmath}
\usepackage{amssymb}
\usepackage{xcolor}
\usepackage{physics}
\usepackage[hidelinks]{hyperref}

\begin{document}

\title{Reentrant Nematicity from Landau Quantization near a Pomeranchuk Instability}

\author{Luís E. Leite Macêdo}
\affiliation{Departamento de F{\'\i}sica Te\'orica,
Universidade do Estado do Rio de Janeiro, Rua S\~ao Francisco Xavier 524, 20550-013
Rio de Janeiro, Brazil}

\author{Daniel G. Barci}
\affiliation{Departamento de F{\'\i}sica Te\'orica,
Universidade do Estado do Rio de Janeiro, Rua S\~ao Francisco Xavier 524, 20550-013
Rio de Janeiro, Brazil}

\date{\today}

\begin{abstract}
Landau quantization can induce nematic order in a two-dimensional Fermi liquid that is isotropic at zero magnetic field. Using a mean-field theory near a quadrupolar Pomeranchuk instability, we show that varying the magnetic field produces alternating nematic and isotropic regions, with transitions that can be either continuous or first order. The resulting nematic state also leaves distinctive signatures in the de Haas--van Alphen oscillations. At fixed chemical potential, the field-dependent deformation of the Fermi surface can produce substantial changes in the oscillation frequency, even when the cyclotron mass varies only modestly. At fixed particle density, conservation of the Fermi-surface area preserves the leading oscillation frequency, while the cyclotron mass and harmonic amplitudes remain sensitive to nematic order. These contrasting behaviors provide experimentally relevant signatures of field-induced nematicity and reveal the importance of thermodynamic constraints in quantum-oscillation measurements.
\end{abstract}

\maketitle
Magnetic fields play a dual role in quantum materials. They are among the most precise probes of electronic structure, but they can also reorganize that structure and stabilize phases that are absent at zero field. This role becomes more apparent  when Landau quantization is well resolved: as the field is varied, discrete Landau levels move through the Fermi energy and can repeatedly change the energetic balance between competing states. Reentrant field-induced phases occur in a wide range of settings, from superconductors and density-wave systems to moiré materials and two-dimensional magnets~\cite{LittleParks1962,GruenbergGunther1968,Tesanovic1989,Naughton1988,SongKoshelev2017,Das2022,Foutty2024}, although their microscopic mechanisms differ. Recent observations of reentrant Landau-level crossings and non-\(1/B\)-periodic oscillations in ZrTe$_5$ further illustrate how strongly a quantized spectrum can evolve with magnetic field~\cite{KaufmannRibeiro2026}.

Quantum oscillations have historically been viewed from a complementary perspective. In the de Haas--van Alphen (dHvA) effect, successive Landau-level crossings produce magnetization oscillations periodic in inverse field, whose frequency, thermal damping, and harmonic content encode information about the Fermi surface and the quasiparticle mass~\cite{Onsager1952,LifshitzKosevich1956,Shoenberg1984}. This spectroscopic role extends well beyond conventional metals: quantum oscillations remain observable in strongly correlated regimes, including systems close to quantum criticality and symmetry-breaking instabilities. In Sr$_3$Ru$_2$O$_7$, for example, dHvA oscillations remain observable in and around the field-induced anisotropic regime near metamagnetism~\cite{Borzi2007,Mercure2009,Mercure2010}. In such situations, Landau quantization need not act only as a passive probe of a preexisting electronic state; sufficiently close to an instability, it can participate directly in selecting that state.

Electronic nematicity provides a natural setting to investigate how Landau quantization influences the stability of electronic order. A nematic metal spontaneously breaks rotational symmetry while preserving translations, so that an initially symmetric electronic fluid develops a preferred spatial direction and its Fermi surface becomes distorted~\cite{OgKiFr2001,Fradkin2010}. Nematic and closely related electronic anisotropies occur in quantum Hall systems, ruthenates, cuprates, and iron-based superconductors~\cite{Fradkin2010,Fernandes2014}. A metal close to a nematic instability is especially sensitive to changes in its low-energy spectrum, and successive Landau-level passages can therefore modulate the balance between the isotropic and symmetry-broken states. Within Fermi-liquid theory, the minimal continuum realization of this physics is a quadrupolar Pomeranchuk instability~\cite{Pomeranchuk1959,OgKiFr2001}. Previous work addressed the perturbative effect of a weak magnetic field on this instability and its collective dynamics~\cite{Barci-Reyes-2013,BarciFradkinRibeiro2018}.

Here we retain the full Landau-level thermodynamics within a self-consistent  mean-field theory and determine the phase structure by global minimization of the thermodynamic potential. We find that Landau quantization generates a sequence of reentrant nematic regions as the magnetic field is swept, even when the zero-field state is isotropic. The reentrance arises because the quantized spectrum modulates, with different phases and amplitudes, both the tendency toward nematic order and the nonlinear energetics that stabilize it. Once nematic order develops, the deformation also modifies the quantum oscillations. At fixed chemical potential, the oscillation phase can change substantially even when the cyclotron mass varies only slightly. At fixed particle density, by contrast, conservation of the Fermi-surface area preserves the leading oscillation frequency.
Landau quantization can thus influence the formation of nematic order, while the resulting Fermi-surface deformation leaves distinct signatures in the quantum oscillations.

\textit{Model and Landau quantization.—}
We use a minimal continuum model of a two-dimensional Fermi liquid near a
quadrupolar Pomeranchuk instability, with spinless fermions and an
attractive quadrupolar interaction,
\begin{equation}
H=\int d^2x\,\psi^\dagger \varepsilon({\bf D})\psi
+\frac{f_2}{4}\int d^2x\,{\rm Tr}\,\hat Q^2,
\label{eq:H}
\end{equation}
where $f_2<0$ is the quadrupolar coupling constant, $\varepsilon({\bf D})$ is a single-band fermionic dispersion, the covariant derivative 
${\bf D}=-i\boldsymbol\nabla+{\bf A}$ with
$\boldsymbol\nabla\times{\bf A}=B\hat{\bf z}$. $B>0$ is the intensity of a constant magnetic field perpendicular to the plane.  The quadrupolar density $\hat Q$
is a symmetric trace-less tensor, whose components are
\[
Q_{ij}=
\psi^\dagger
\left[
\frac{1}{2}\{D_i,D_j\}
-\frac{\delta_{ij}}{2}{\bf D}^2
\right]\psi .
\]
with the latin index $i,j$ taking the values $1$ or $2$ . We use units $\hbar=k_B=e=1$.
We omit spin to isolate the orbital feedback of Landau
quantization. The covariant derivatives make the quadrupolar density gauge
invariant, while the perpendicular field preserves in-plane rotations and
therefore does not explicitly select a nematic director. At $B=0$,
Eq.~(\ref{eq:H}) reduces to the standard quadrupolar Fermi liquid
\cite{OgKiFr2001,BaOx2003,Lawler2006}.

We decouple the quadrupolar interaction through a Hubbard--Stratonovich transformation and integrate out the fermions. We then use a static,
spatially uniform mean-field approximation. The nematic state is then
described by a dimensionless order-parameter amplitude $q$, which
measures the strength of the quadrupolar distortion of the electronic
dispersion, 
\begin{equation}
q^2\equiv \frac{m^2|f_2|}{2}{\rm Tr}\langle\hat Q^2\rangle_{MF}
\end{equation}
where  $\langle\ldots\rangle_{MF}$ stands for  mean field expectation value. 
Thus,  $q=0$ represents an isotropic, whereas $q\neq0$ corresponds to a
nematic state. 

For a parabolic band,
$\varepsilon({\bf D})={\bf D}^2/(2m)$, the corresponding Landau problem is
exactly reducible to the isotropic one. The discrete energy spectrum is
\begin{equation}
E_\lambda(q)=
\left(\lambda+\frac12\right)\omega_c\Delta(q),
\label{eq:LL}
\end{equation}
where $\omega_c=B/m$ and $\Delta(q)=\sqrt{1-q^2}$, with $|q|<1$.
Nematicity therefore changes the Landau-level spacing,
equivalently the cyclotron mass $m_c=m/\Delta(q)$, while leaving the
degeneracy per physical area, $
{\cal D}_B=B/2\pi$, unchanged. Since the Landau-level energies
depend on $q$, the oscillatory part of the thermodynamic potential also
depends on the nematic order parameter and feeds back into its
self-consistent determination.

\textit{Self-consistent thermodynamics.—}
At fixed chemical potential, $\mu$, the uniform mean-field
grand potential can be written as
\begin{equation}
\Omega_{\rm MF}(q)=
\Omega_{f,{\rm par}}(q)
+\frac{a}{2|F_2|}q^2
+\delta\Omega_{\rm stab}(q).
\label{eq:OmegaMF}
\end{equation}
Here $a=m\mu^2/(4\pi)$ is the natural energy-density scale and
$F_2=m^2f_2a<0$ is the dimensionless quadrupolar coupling.

The term $\Omega_{f,{\rm par}}$ is the parabolic-band fermionic contribution,
including its smooth and oscillatory parts, and the quadratic term is the
interaction cost. The finite-amplitude potential is stabilized by an
irrelevant radial-curvature correction $\delta\Omega_{\rm stab}$~\cite{OgKiFr2001,BaOx2003}.
Its microscopic form and the smooth thermodynamic terms are given in the
Supplemental Material; the oscillatory contribution is
\begin{align}
\Omega_{\rm osc}(q)=&\;
a\left(\frac{\omega_c}{\mu}\right)^2
\Delta(q)
\sum_{n=1}^{\infty}
\frac{(-1)^n}{\pi^2n^2}
R_T\!\left(\frac{\chi_n}{\Delta(q)}\right)
\nonumber\\
&\times
\cos\!\left[
\frac{2\pi n\mu}{\omega_c\Delta(q)}
\right],
\label{eq:Omegaosc}
\end{align}
where $R_T(x)=x/\sinh x$ is the Lifshitz--Kosevich (LK) thermal damping factor and $\chi_n=2\pi^2nT/\omega_c$.
Equation~(\ref{eq:Omegaosc}) exposes the feedback mechanism:
the nematic amplitude enters not only the LK damping $R_T$ but
also the rapidly varying dHvA phase.
The grand potential in Eq.~(\ref{eq:OmegaMF}) is evaluated in the
degenerate regime $T/\mu\ll1$ and the weak-field regime
$\omega_c/\mu\ll1$, with no expansion in $T/\omega_c$.

Close to the isotropic state,
\begin{equation}
\frac{\Omega_{\rm MF}(q)-\Omega_{\rm MF}(0)}{a}
=
\frac{r}{2}q^2+\frac{u}{4}q^4+\cdots ,
\label{eq:Landau}
\end{equation}
where, at $B=T=0$, $r=(1-|F_2|)/|F_2|$. The coefficient $r$ controls the
tendency toward nematic order, while $u$ controls the nonlinear stability
of the finite-amplitude state. The
zero-field mean-field transition is continuous at $F_{2,c}=-1$: the state
is isotropic for $-1<F_2<0$ and nematic for $F_2<-1$~\cite{OgKiFr2001,Lawler2006}.

In the thermally damped first-harmonic regime at finite $B$, the quadratic and quartic coefficients can be written as
$r=\bar r+r_{\rm osc}$ and $u=\bar u+u_{\rm osc}$, with bars denoting
the smooth contributions. To leading order in $\omega_c/\mu$, 
\begin{align}
r_{\rm osc}&\simeq
\frac{2}{\pi}\frac{\omega_c}{\mu}
R_T(\chi_1)\sin\theta,
\label{eq:r}\\
u_{\rm osc}&\simeq 2R_T(\chi_1)\cos\theta,
\label{eq:ru}
\end{align}
where 
$\theta=2\pi\mu/\omega_c$.
The quadratic and quartic sectors are approximately in quadrature, but their
oscillatory amplitudes are parametrically different: the oscillatory
shift of the instability is $O(\omega_c/\mu)$, whereas the leading
nonlinear modulation is $O(1)$. Near a Landau-level crossing, the leading
shift of $r$ vanishes where the reduction of $u$ is maximal.
Thus, the same Landau quantization that produces dHvA oscillations enters directly into the stability criteria of the nematic phase, changing not only where nematicity appears but also the nonlinear structure of the transition.

At low temperatures, the local expansion becomes nonuniform near Landau-level
crossings. To obtain the equilibrium phase diagram, we therefore perform a global
minimization of the full finite-temperature potential, retaining the
$q$ dependence of both the phase and the thermal factor in
Eq.~(\ref{eq:Omegaosc}).
Continuous transitions occur when the minimum evolves continuously from
$q=0$; first-order transitions exchange stability between distinct
isotropic and nematic minima. The Supplemental Material gives the
minimization procedure and convergence checks; the companion
paper~\cite{MacedoBarciCompanion} provides the broader thermodynamic and
robustness analysis.

\begin{figure}[tbp]
    \centering
    \includegraphics[width=\columnwidth]{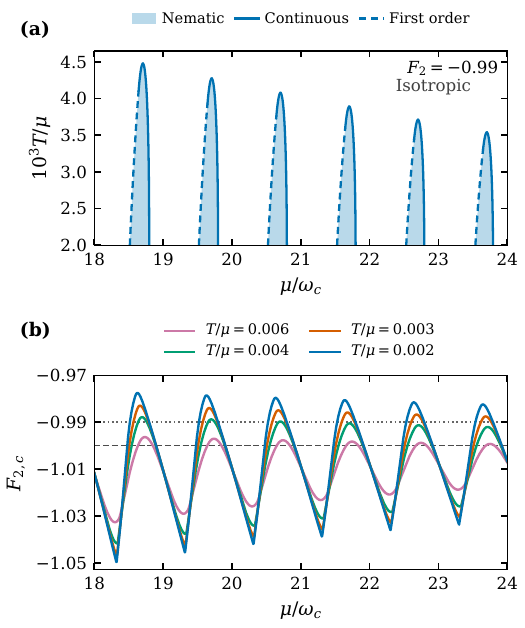}
    \caption{Reentrant nematic regions at fixed chemical potential.
    (a) Phase diagram for $F_2=-0.99$, on the zero-field isotropic side.
    Filled regions are nematic while the white regions are isotropic.
    Solid and dashed boundaries denote continuous and first-order
    transitions. (b) Global critical coupling versus inverse field at
    $T/\mu=0.006,0.004,0.003,$ and $0.002$. The boundary crosses the
    zero-field threshold $F_2=-1$ (dashed line), inducing order for
    $F_2>-1$. The dotted line marks the cut $F_2=-0.99$ in (a).}
    \label{fig:phase_diagram}
\end{figure}

\textit{Landau-quantization-driven nematicity.—}
In  Fig.~\ref{fig:phase_diagram}(a) we show the phase diagram in the 
$(T/\mu,\mu/\omega_c)$ plane, for $F_2=-0.99$.
Note that,   for this value of $F_2$, the zero-field state is isotropic. Global minimization
nevertheless yields alternating nematic and isotropic intervals as the
magnetic field is swept. The nematic
lobes in the $(T/\mu,\mu/\omega_c)$ plane demonstrate repeated
transitions between isotropic and nematic states induced by the quantized spectrum.
In Fig.~\ref{fig:phase_diagram}(b), we show the critical quadrupolar interaction $F_{2,c}$ as a function of 
$\mu/\omega_c$ for different values of temperature.   The critical boundary crosses into $F_2>-1$, so the field creates order
where the zero-field system remains isotropic.
Lowering $T/\mu$ weakens Lifshitz--Kosevich damping and increases these
excursions. The discrete spectrum thus repeatedly reverses the energetic
balance between isotropic and nematic states.

\begin{figure}[tbp]
    \centering
    \includegraphics[width=\columnwidth]{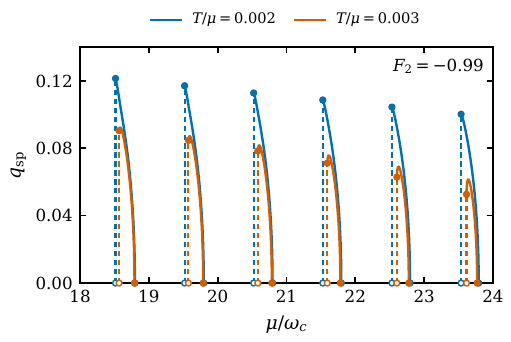}
    \caption{Repeated appearance and disappearance of the equilibrium
    nematic amplitude $q_{\rm sp}$ at $F_2=-0.99$, for $T/\mu=0.002$
    and $0.003$. Each displayed lobe has a finite jump on one side and
    continuous vanishing on the other. Dashed vertical lines join coexisting
    isotropic (open markers) and nematic (filled markers) amplitudes;
    solid curves trace the nematic branch.}
    \label{fig:qsp}
\end{figure}

In Fig.~\ref{fig:qsp}, we show the order parameter $q_{\rm sp}(T,B)$ at the saddle-point (mean-field) approximation as a function of $\mu/\omega_c$ for  two values of the temperature along the same $F_2=-0.99$ cut.
The order parameter amplitude repeatedly appears and
disappears at both temperatures. Each displayed lobe is bounded
on one side by a finite jump of $q_{\rm sp}$ and on the other by a
continuous vanishing of the order parameter. Successive Landau-level
passages can therefore periodically change not only the existence of the
ordered state but also the character of its boundary. This behavior is the
global, finite-amplitude counterpart of the strongly oscillating quartic
sector in Eq.~(\ref{eq:ru}).

The reentrant structure and its first-order segments are robust with respect to
the model details considered in Ref.~\cite{MacedoBarciCompanion}. However, the precise coexistence fields, jump amplitudes,
and the locations  of endpoints are more sensitive to short-distance stabilization.

The calculated boundaries refer to the static uniform mean-field sector.
Of course, in a strictly rotationally invariant two-dimensional model,
orientational fluctuations replace conventional finite-temperature
long-range nematic order by Berezinskii--Kosterlitz--Thouless
physics~\cite{MeWa1966,KoTh1973,BaSt2007-1,BaSt2009}.
However, lattice anisotropy can reduce the symmetry to a discrete one and permit a
true finite-temperature transition~\cite{KKC-2003}.

\textit{Quantum-oscillation signatures.—}
The reentrant nematic state also modifies the quantum oscillations.
At fixed chemical potential, the phase of the $\ell$th harmonic and the
cyclotron mass are
\begin{equation}
\Phi_{\mu,\ell}
=
\frac{2\pi\ell\mu}{\omega_c\Delta(q_{\rm sp})},
\qquad
m_c=\frac{m}{\Delta(q_{\rm sp})}.
\label{eq:phase_mu}
\end{equation}
For $q_{\rm sp}\ll1$, the mass and phase shifts satisfy
\begin{equation}
\frac{\delta m_c}{m}
\simeq
\frac{q_{\rm sp}^2}{2},
\qquad
\frac{\delta\Phi_{\mu,\ell}}{2\pi\ell}
\simeq
\frac{\mu}{\omega_c}\frac{q_{\rm sp}^2}{2}.
\label{eq:phase_amplification}
\end{equation}
The normalized phase displacement in Eq.~(\ref{eq:phase_amplification})
is enhanced by $\mu/\omega_c$ relative to the fractional mass change.
Thus a modest deformation can appreciably shift oscillation extrema.
Equation~(\ref{eq:phase_amplification}) describes a field-dependent phase
shift rather than a direct frequency shift. Because $q_{\rm sp}$ varies with
field, this phase modulation can nevertheless change the spacing between
successive oscillation extrema in $1/B$.

The strength of this phase feedback depends crucially on the thermodynamic
constraint. At fixed density, the Fermi-surface area remains fixed, even in the nematic phase. Consequently, the Onsager frequency is not renormalized by the Fermi-surface deformation. For the parabolic saddle, this cancellation can be evaluated exactly. The chemical potential varies as 
$\bar\mu(q)=\mu_0\Delta(q)$, canceling the deformation of the rapidly
varying phase,
\begin{equation}
\Phi_{N,\ell}
=
\frac{2\pi\ell\mu_0}{\omega_c},
\qquad
\nu_{B,N}=m\mu_0,
\label{eq:phase_N}
\end{equation}
where $\mu_0$ is the isotropic zero-field chemical potential at the
specified density and $\nu_{B,N}$ is the corresponding Onsager frequency.
The cyclotron mass still changes as $m_c=m/\Delta(q)$, and thermal
damping and oscillation amplitudes can evolve even while the leading
Onsager phase is locked by area conservation.
Fixed chemical potential and fixed density should therefore be viewed as limiting cases
rather than mutually exclusive experimental situations. In a multiband
metal, additional Fermi-surface sheets can act as particle reservoirs and
partially pin the chemical potential, while the total density remains fixed
\cite{ItskovskyManivVagner2000}. The response of an individual oscillating
pocket can then interpolate between the grand-canonical and
canonical limits.

Sr$_3$Ru$_2$O$_7$ provides an interesting experimental context for our results. Strongly field-dependent dHvA frequencies have been observed near its field-induced anisotropic phase, adjacent to the metamagnetic region, without comparably large changes in the cyclotron masses of the reliably resolved branches. Chemical-potential variations also contribute to the observed frequency evolution~\cite{Borzi2007,Mercure2009,Mercure2010}. In a multiband metal, some bands may act as particle reservoirs, reducing these variations and allowing the system to approach the fixed-chemical-potential regime. Although a quantitative description of Sr$_3$Ru$_2$O$_7$ requires accounting for spin, multiband structure, magnetic breakdown, and metamagnetism, our results suggest that field-dependent nematic order may contribute to the unusual evolution of its quantum oscillations.

\textit{Conclusions.—}
Landau quantization can repeatedly induce and suppress nematic order in a two-dimensional Fermi liquid that is isotropic at zero field but close to a Pomeranchuk instability. This behavior arises because the magnetic field modifies the quadratic and higher-order terms of the free energy in different ways. The resulting nematic deformation also changes the dHvA oscillations, with distinct consequences at fixed chemical potential and fixed particle density. Together, these results establish a connection between magnetic-field-induced electronic order and quantum oscillations, offering a useful perspective for investigating correlated metals near electronic instabilities.

\textit{Acknowledgments.—}
We acknowledge partial financial support from the Brazilian agencies {\em Funda\c c\~ao Carlos Chagas Filho de Amparo \`a Pesquisa do Estado do Rio de Janeiro} (FAPERJ), {\em Conselho Nacional de Desenvolvimento Cient\'\i fico e Tecnol\'ogico} (CNPq), and {\em Coordena\c c\~ao de Aperfei\c coamento de Pessoal de N\'\i vel Superior} (CAPES, Finance Code 001).
We acknowledge the support of the INCT project Advanced Quantum Materials, involving the Brazilian agencies CNPq (Proc. 408766/2024-7), FAPESP (Proc. 2025/27091-3), and CAPES.
L.E.L.M. acknowledges a CAPES PhD fellowship.

\clearpage
\setcounter{equation}{0}
\renewcommand{\theequation}{S\arabic{equation}}
\setcounter{figure}{0}
\renewcommand{\thefigure}{S\arabic{figure}}
\setcounter{table}{0}
\renewcommand{\thetable}{S\arabic{table}}
\setcounter{section}{0}

\begin{center}
{\large\bfseries Supplemental Material for ``Reentrant Nematicity from Landau Quantization near a Pomeranchuk Instability''\par}
\end{center}

\section{Global minimization of the mean-field thermodynamic potential}

This Supplemental Material describes the numerical procedure used to obtain the equilibrium phase diagrams and order-parameter profiles discussed in the Letter. The central point is that the phase boundaries are determined from the global minimum of the finite-temperature mean-field thermodynamic potential, rather than from a truncation of its local Landau expansion. This distinction becomes important at low temperature, where the rapidly varying de Haas--van Alphen phase can make the expansion in the nematic amplitude nonuniform near individual Landau-level crossings.

We work at fixed chemical potential and define
\begin{equation}
a=\frac{m\mu^2}{4\pi},\qquad
\epsilon_B=\frac{\omega_c}{\mu},\qquad
\Delta(q)=\sqrt{1-q^2},
\end{equation}
where $0\le q<1$ is the dimensionless amplitude of the uniform nematic deformation, and $F_2=m^2 a f_2<0$ is the dimensionless attractive quadrupolar coupling. Following the mean-field construction developed in detail in the companion paper~\cite{SM:MacedoBarciCompanion}, the uniform grand potential is

\begin{align}
\Omega_{\rm MF}(q)&=\Omega_{f,{\rm par}}(q)
+\frac{a}{2|F_2|}q^2+\delta\Omega_\gamma(q)
\label{S:OmegaMF}
\end{align}
The fermionic part, $\Omega_{f,{\rm par}}(q)$, contains the smooth zero-field, nonoscillatory magnetic, and oscillatory contributions of the parabolic spectrum. The quadratic term proportional to $q^2$ is the Hubbard--Stratonovich interaction cost, and $\delta\Omega_\gamma$ is the stabilizing radial-curvature contribution (denoted $\delta\Omega_{\rm stab}$ in the Letter).

The individual fermionic terms can be written as
\begin{align}
\Omega_{f,{\rm par}}(q)&=\Omega_f^0(q)
+\Omega_f^{\rm non}(q)+\Omega_f^{\rm osc}(q).
\end{align}
Specifically,
\begin{align}
\Omega_f^0(q)&=-\frac{ a A_T}{\Delta(q)},
\label{S:Omegaf0}\\
\Omega_f^{\rm non}(q)&=\frac{a\epsilon_B^2}{12}
\Delta(q),
\label{S:Omegafnon}\\
\Omega_f^{\rm osc}(q)&=a\epsilon_B^2 S[\Delta(q)],
\label{S:Omegafosc}
\end{align}
where
\begin{equation}
A_T=
1+\frac{\pi^2}{3}\left(\frac{T}{\mu}\right)^2
\end{equation}
and
\begin{equation}
\label{eq:sm-oscillatory-sum}
S(\Delta)=
\Delta
\sum_{n=1}^{n_{\rm max}}
\frac{(-1)^n}{\pi^2n^2}
R_T\!\left(\frac{\chi_n}{\Delta}\right)
\cos\!\left(\frac{\theta_n}{\Delta}\right).
\end{equation}
Here,
\begin{equation}
R_T(x)=\frac{x}{\sinh x},
\qquad
\chi_n=\frac{2\pi^2nT}{\omega_c},
\qquad
\theta_n=\frac{2\pi n\mu}{\omega_c}.
\end{equation}
The function $R_T$ is the standard Lifshitz--Kosevich thermal reduction
factor~\cite{SM:LifshitzKosevich1956,SM:Shoenberg1984}.

The equilibrium state is obtained by minimizing the dimensionless potential difference
\begin{equation}
V_\mu(q)\equiv
\frac{\Omega_{\rm MF}(q)-\Omega_{\rm MF}(0)}{a}.
\end{equation}

Using Eqs. (\ref{S:OmegaMF})-(\ref{S:Omegafosc}), the potential used in the numerical calculation takes the form
\begin{align}
V_\mu(q)=&
\frac{q^2}{2|F_2|}
-A_T\left(\frac{1}{\Delta}-1\right)
+\frac{\epsilon_B^2}{12}(\Delta-1)
\nonumber\\
&+\epsilon_B^2\left[S(\Delta)-S(1)\right]
+\delta\omega_\gamma(q),
\label{eq:sm-full-potential}
\end{align}
where $\delta\omega_\gamma(q)=[\delta\Omega_\gamma(q)-\delta\Omega_\gamma(0)]/a$ is the dimensionless stabilizing contribution, whose finite-amplitude continuation is constructed in the next section.
Neither the rapidly varying phase $\theta_n/\Delta(q)$ nor the thermal factor $R_T[\chi_n/\Delta(q)]$ is expanded in powers of $q$. Thus the numerical potential retains the full deformation dependence generated by the parabolic Landau spectrum.

The local expansion
\begin{equation}
V_\mu(q)=
\frac{r_\mu(T,B)}{2}q^2+
\frac{u_\mu(T,B)}{4}q^4+\cdots
\end{equation}
is used only to characterize the stability of the isotropic saddle and to interpret the origin of the oscillatory phase boundary; it is not extrapolated into the finite-amplitude regime when determining the equilibrium state.
Near a Landau-level crossing, the local expansion requires $\mu q^2\ll T$; its domain therefore shrinks as $T\to0$, anticipating the nonanalytic zero-temperature crossing. This restriction motivates using the full finite-temperature potential in the global minimization.

\section{Stabilizing Fermi-surface correction}

The strictly parabolic dispersion does not stabilize the nematic state beyond the local isotropic instability. We therefore include the gauge-covariant radial correction
\begin{equation}
\label{eq:sm-radial-operator}
\delta\varepsilon_\gamma({\bf D})=
\frac{\gamma}{8mk_F^4}
\left({\bf D}^2-k_F^2\right)^3,
\qquad
k_F^2=2m\mu.
\end{equation}
Near the zero-field Fermi surface, this reproduces the cubic radial-curvature term
\begin{equation}
\varepsilon(k)-\mu
\simeq
v_F(k-k_F)
+\frac{\gamma}{mk_F}(k-k_F)^3+\cdots,
\end{equation}
which modifies the nonlinear stability of the nematic state without shifting the zero-field quadratic Pomeranchuk threshold.
At $B=T=0$, this correction gives $u=3(\gamma-8)/16$, so $\gamma>8$
ensures a positive local quartic coefficient while leaving $F_{2,c}=-1$
unchanged~\cite{SM:MacedoBarciCompanion}.
Nonlinear radial corrections of this type stabilize continuum theories
of nematic Fermi fluids beyond the quadratic Pomeranchuk
instability~\cite{SM:OgKiFr2001,SM:BaOx2003}.

At leading order in perturbation theory, the correction in Eq.~(\ref{eq:sm-radial-operator}) shifts each Landau-level energy by
\begin{align}
\delta E_\lambda(q)=
\frac{\gamma}{8mk_F^4}
\Big[
&\langle({\bf D}^2)^3\rangle_\lambda
-3k_F^2\langle({\bf D}^2)^2\rangle_\lambda
\nonumber\\
&+3k_F^4\langle{\bf D}^2\rangle_\lambda
-k_F^6
\Big].
\end{align}
Here $\langle\ldots\rangle_\lambda$ denotes the expectation value in the
anisotropic Landau-level basis, with unperturbed energies
\begin{equation}
E_\lambda^{(0)}(q)=
\left(\lambda+\frac12\right)\omega_c\Delta(q).
\end{equation}
The degeneracy is ${\cal D}_B=B/2\pi$.
A direct calculation gives
\begin{align}
\langle{\bf D}^2\rangle_\lambda
&=
\frac{Bs_\lambda}{\Delta},
\\
\langle({\bf D}^2)^2\rangle_\lambda
&=
B^2
\left[
s_\lambda^2+
\frac{3q^2}{2(1-q^2)}
(s_\lambda^2+1)
\right],
\\
\langle({\bf D}^2)^3\rangle_\lambda
&=
\frac{B^3s_\lambda}
{2(1-q^2)^{3/2}}
\left[
2s_\lambda^2+
q^2(3s_\lambda^2+25)
\right].
\end{align}
Here $s_\lambda=2\lambda+1$.

The regulated correction to the dimensionless fermionic grand potential is
\begin{equation}
G_W(q)=
\frac{{\cal D}_B}{a}
\sum_{\lambda\ge0}
W_{\Lambda,p}(\xi_\lambda)
n_F\!\left[E_\lambda^{(0)}(q)-\mu\right]
\delta E_\lambda(q),
\end{equation}
where $n_F(\varepsilon)=[e^{\varepsilon/T}+1]^{-1}$
denotes the Fermi--Dirac distribution. We introduce the following smooth regulator to restrict the sum to an energy shell around the Fermi surface:
\begin{equation}
W_{\Lambda,p}(\xi)
=
\exp\left[
-\left|\frac{\xi}{\Lambda}\right|^p
\right],
\qquad
\xi_\lambda=
\left(\lambda+\frac12\right)\omega_c-\mu.
\end{equation}
The regulator is referenced to the isotropic spectrum and is independent of $q$; the cutoff therefore does not move with the nematic deformation.

We define the correction to the dimensionless grand potential as
\begin{align}
\delta\omega_\gamma(q)=&
G_W(q)-G_W(0)
+
\left(
\frac{\delta r_\gamma}{2}-g_2
\right)q^2
\nonumber\\
&+
\left(
\frac{\delta u_\gamma}{4}-g_4
\right)q^4
+c_6q^6+c_8q^8.
\end{align}
Here the counterterms preserve the prescribed quadratic and quartic coefficients of the low-energy theory while allowing the potential to be continued to finite $q$. The analytic corrections $\delta r_\gamma$ and $\delta u_\gamma$ entering this equation are derived explicitly in the companion paper~\cite{SM:MacedoBarciCompanion}.
The coefficients $c_6$ and $c_8$ are set to zero in the reference calculation and varied only in robustness tests of the numerical results.
The coefficients $g_2$ and $g_4$ are obtained from the small-$q$ expansion
\begin{equation}
G_W(q)=G_W(0)+g_2q^2+g_4q^4+\cdots
\end{equation}

\section{Global numerical minimization}

For each set of parameters $(T,B,F_2)$, the interval
$
0\le q<1
$
is searched for local minima of $V_\mu(q)$. The potential is first sampled on a nonuniform grid designed to resolve the small-$q$ region while retaining coverage of the entire positive-definite domain. All candidate local minima detected on this grid are bracketed and subsequently refined by one-dimensional minimization. The resulting values of the thermodynamic potential are then compared with each other and with the isotropic value
\begin{equation}
V_\mu(0)=0.
\end{equation}

The equilibrium order parameter is therefore
\begin{equation}
q_{\rm sp}(T,B,F_2)
=
\underset{0\le q<1}{\operatorname{argmin}}\,
V_\mu(q).
\end{equation}
The minimization is performed independently at each parameter point. In particular, the calculation does not determine the equilibrium state by continuously following one local minimum as the magnetic field is swept. Metastable solutions can therefore be identified without being confused with the equilibrium branch.

The same global comparison can be formulated directly for the phase boundary. Writing
\begin{equation}
V_\mu(q)=
\frac{q^2}{2|F_2|}+K(q),
\end{equation}
where $K(q)$ is independent of $F_2$, the critical interaction satisfies
\begin{equation}
\label{eq:sm-global-boundary}
\frac{1}{|F_{2,c}|}
=
\max
\left\{
{\cal X}_2(T,B),
\;
\sup_{0<q<1}
\left[
-\frac{2K(q)}{q^2}
\right]
\right\}.
\end{equation}
Here ${\cal X}_2(T,B)$ is the local quadratic susceptibility entering
\begin{equation}
r_\mu(T,B)=\frac{1}{|F_2|}-{\cal X}_2(T,B).
\end{equation}

If the maximum in Eq.~\eqref{eq:sm-global-boundary} is attained only in the limit $q\rightarrow0$, the equilibrium boundary is continuous and coincides with
\begin{equation}
r_\mu(T,B)=0,
\qquad
u_\mu(T,B)>0,
\end{equation}
provided no competing finite-amplitude minimum is present.

A first-order transition instead occurs when a finite-$q$ minimum becomes degenerate with the isotropic state,
\begin{equation}
V_\mu(q_N)=V_\mu(0)=0,
\qquad
q_N>0.
\end{equation}
Thus the first-order boundary is determined by equality of the competing thermodynamic potentials, rather than by a spinodal or the local condition $r_\mu=0$. This global criterion also detects finite-amplitude preemption when the local quartic coefficient remains positive.

This distinction is important near a Landau-level crossing. The local quadratic coefficient controls the loss of stability of the isotropic saddle, while the nonlinear part of the potential can vary much more strongly with magnetic field. Consequently, the local instability and the equilibrium transition need not coincide.

\begin{figure}[t]
\centering
\includegraphics[width=\columnwidth]{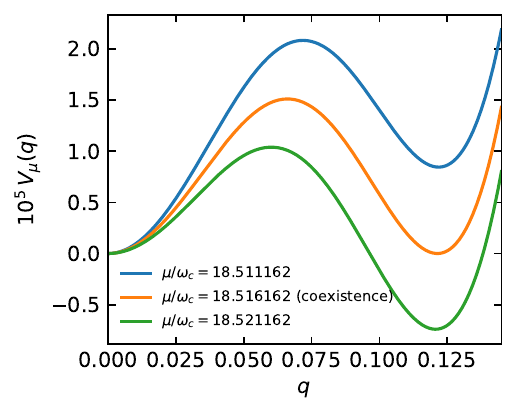}
\caption{Representative full-potential profiles across a first-order
nematic boundary for $T/\mu=0.002$ and $F_2=-0.99$. The curves show
the evolution from a globally stable isotropic state with a metastable
finite-$q$ minimum, through coexistence of degenerate isotropic and
nematic minima, to a globally stable nematic state.}
\label{fig:potential_profiles}
\end{figure}
For the representative boundary shown in Fig.~\ref{fig:potential_profiles},
the transition is weakly first order, with a relatively small
discontinuity of the nematic order parameter and a shallow free-energy
barrier compared with the microscopic scale
$a=m\mu^2/(4\pi)$. Its thermodynamic classification is nevertheless
unambiguous: at coexistence the isotropic minimum at $q=0$ and a
distinct finite-$q$ minimum have equal free energy. The weakly
first-order character refers to this representative effective-potential
realization; the existence of finite-amplitude coexistence is more
robust than the precise values of the jump and barrier height.

\section{Numerical convergence and robustness}

The oscillatory sum in Eq.~\eqref{eq:sm-oscillatory-sum} is truncated at a harmonic number $n_{\rm max}$ chosen by direct convergence tests. As a representative benchmark, for
\begin{equation}
T/\mu=0.006,\qquad
\gamma=10,\qquad
F_2=-1.0078,
\end{equation}
the coexistence inverse field obtained with $n_{\rm max}=3,5,10$ is, respectively,
\begin{equation}
18.54675125,\qquad
18.54657908,\qquad
18.54657836.
\end{equation}
Calculations with $n_{\rm max}=20$ and $40$ agree with the $n_{\rm max}=10$ result at the displayed precision, while the coexistence jump changes by less than $10^{-6}$ between $n_{\rm max}=10$ and $20$. The harmonic cutoff is increased until both the phase boundary and the equilibrium order parameter are numerically stable.

For the reference finite-amplitude continuation we use
\begin{equation}
\gamma=10,\qquad
\frac{\Lambda}{\mu}=0.3,\qquad
p=4,\qquad
c_6=c_8=0.
\end{equation}
The robustness of the first-order structure was tested by replacing
$
(\Lambda/\mu,p)=(0.3,4)
$
with
$
(0.2,2),
(0.5,4)$ and
$(0.3,6)$.

For the representative coexistence point specified above, these variations preserve the first-order transition and give
\begin{equation}
18.54618
\lesssim
\frac{\mu}{\omega_c}
\lesssim
18.54673,
\end{equation}
and
\begin{equation}
0.09804
\lesssim
q_{\rm jump}
\lesssim
0.10005.
\end{equation}
No lower large-$q$ minimum is detected by the global search.

Higher-order variations provide a complementary test. Since two effective potentials with the same prescribed $r_\mu$ and $u_\mu$ first differ at order $q^6$, the detailed classification of some endpoints is necessarily more sensitive to short-distance stabilization than the existence of the finite-amplitude coexistence regions themselves. Variations of $c_6$ and $c_8$ modify the higher-order coefficients without changing the matched quadratic and quartic terms. The tested first-order segments persist under these variations, although the location and classification of individual endpoints may change.

The regulated level sums are truncated only after the regulator exponent becomes sufficiently large that additional levels are numerically irrelevant. Changing this cutoff over the tested range leaves the representative higher-order checks unchanged at the quoted precision. Independent high-precision differentiation verifies that the matched finite-amplitude potential reproduces the prescribed quadratic and quartic coefficients, while mesh refinement and direct equality-of-potential tests verify the global minimization. At the representative coexistence point, the difference between the two degenerate minima is below $10^{-14}a$.

\section{Range of validity}

The numerical phase diagrams describe the static, spatially uniform mean-field saddle in the degenerate weak-field regime,
\begin{equation}
\frac{T}{\mu}\ll1,
\qquad
\frac{\omega_c}{\mu}\ll1.
\end{equation}
The parabolic contribution is evaluated with its full dependence on the nematic amplitude, so the global minimization is not restricted by a truncation of the rapidly varying dHvA phase in powers of $q$.

The stabilizing dispersion correction, however, remains a low-energy Fermi-surface construction. Its use is controlled by the smallness of the energy and deformation scales over which the irrelevant operator is sampled. Suppressing the numerical factor $1/4$, the parametric low-energy condition is
\begin{equation}
\gamma
\left[
\frac{\max(T,\omega_c)}{\mu}
\right]^2
\ll1,
\end{equation}
together with a small-deformation condition characterized by
\begin{equation}
\gamma q^2\lesssim1,
\end{equation}
with the quantitative results restricted to the more conservative regime $\gamma q^2\ll1$. The regulated continuation introduces the additional diagnostic
\begin{equation}
\eta_\Lambda
\sim
\frac{\gamma}{4}
\left(\frac{\Lambda}{\mu}\right)^2,
\end{equation}
whose influence is assessed by the explicit regulator variations described above.

The global search over $0\le q<1$ should therefore be understood as a test for competing minima of the specified effective potential. The physically interpreted phase boundaries and order-parameter jumps lie in the small-deformation regime, whereas the absence of a lower minimum at large $q$ is used only as a consistency check and is not regarded as a quantitative prediction of the low-energy theory.

\noindent\textit{Data availability.---}
A reproducibility package containing the numerical code and data required
to reproduce the results of this work is publicly available in the Zenodo repository~\cite{SM:MacedoBarciCode}.

\noindent\textit{AI-assisted code development.---}
Parts of the Python code used for the numerical calculations were developed
with assistance from OpenAI ChatGPT (GPT-5.6 Sol). The AI-assisted code was
reviewed by the authors, executed locally, and validated through convergence
tests, analytic limits, and internal consistency checks. The authors take
full responsibility for the numerical implementation and reported results.

\makeatletter
\begingroup\let\label\@gobble\makeatother

\endgroup
\end{document}